# Observation of gigantic orbital mixing conductance

Yuejie Zhang[1*], Jinjun Ding[2*], Tao Liu[3], Xiaofei Yang[1], Taoyuan Ouyang[4†], Shi Chen[1†], Yongqing Peng[4†]

1. School of Integrated Circuit, Huazhong University of Science and Technology, Wuhan 430074, China
2. Google LLC, Mountain View, California 94043, USA
3. National Engineering Research Center of Electromagnetic Radiation Control Materials, University of Electronic Science and Technology of China, Chengdu 610054, China
4. Beijing Research Institute of Telemetry,100076, Beijing, China

Corresponding author：ouyangty@foxmail.com, s_chen@hust.edu.cn, pengyq@brit.com.cn

**Abstract:**

We report that due to the orbital Hall effect, orbital pumping effects can occur in materials with weak spin-orbit coupling. Moreover, there is a positive correlation between the strength of the orbital Hall effect and the size of spin-pumping. During the spin-pumping, with the enhancement of the orbital Hall effect, the resonant absorption of orbital current and the damping of the ferromagnetic layer also increase. Especially, when the thickness of Ti reaches 60 nm, the orbital-mixing conductance of Ti/Co is an order of magnitude higher than spin-mixing conductance of heavy metal/Co, reaching $474.1 \pm 3 \times 10^{18}\, m^{-2}$. The results indicate that the orbital current is more easily transmitted across the interface.

# 1. Introduction

In non-magnetic metal/ferromagnetic heterostructures, upon the injection of charge current, the non-magnetic metal can generate a polarized current to control the magnetization of magnetic material. The polarized current primarily includes spin-polarized current and orbital-polarized current[1], as shown in Figure 1. While both types of polarized currents originate from non-magnetic materials, the methods of manipulating the magnetization of the magnetic layer are entirely different. Spin-polarized current, upon entering the magnetic material, can directly undergo spin-torque transfer. Materials with strong spin-orbit coupling, such as heavy metals[2] (Pt[3][4][5], Ta[6], W[7], etc.), topological materials (Sn[8][9], BiSb[10][11], $Bi_2Se_3$[12]), and Rashba materials (Ag[13], α-GeTe[14]), showcase polarized currents primarily dominated by spin-polarized currents. The efficiency of spin-torque transfer is primarily determined by the non-magnetic layer's spin Hall angle ($\theta_{SH}$), spin-mixing conductance ($G_{\text{eff}}^{\uparrow\downarrow}$), and the spin transport property ($G_{NM}$) of the non-magnetic material[15]. Therefore, the damping-like torque ($\xi_{DL}$) relationship is given by $\xi_{DL} = \theta_{SH}(2G_{\text{eff}}^{\uparrow\downarrow}/G_{NM})$[1]. However, orbital current entering the magnetic material relies on the magnetic material's spin-orbit coupling (SOC) to first convert it into spin-polarized current before undergoing spin-torque transfer[16][17]. This results in the orbital torque efficiency ($\xi_{DL}$) of the non-magnetic metal being strictly dependent on the SOC strength of the magnetic layer as $\xi_{DL} = C_{FM}\theta_{OH}(2G_{\text{eff}}^{\uparrow\downarrow}/G_{NM})$Error! Reference source not found., where $C_{FM}$ represents the conversion efficiency from orbital current to spin current in the magnetic layer.

Optimizing the spin-mixing conductance is one of the crucial methods to enhance the damping-like torque (DLT)[18]. Spin pumping is a suitable method for studying spin-mixing conductance for non-magnetic metals with strong SOC[19][20]. However, the recently observed inverse orbital Hall effect (IOHE)[21] provides a new approach for studying Spin Pumping[22][23][24] and the spin-mixing conductance of orbital currents. For non-magnetic metals with weak SOC, spin pumping generates a spin current in the magnetic layer, which is subsequently converted into an orbital current within the magnetic layer. Subsequently, as the orbital current enters the non-magnetic metal, the inverse orbital Hall effect occurs. The strength of these effects usually depends on the orbital Hall effect (OHE) in the nonmagnetic layer.

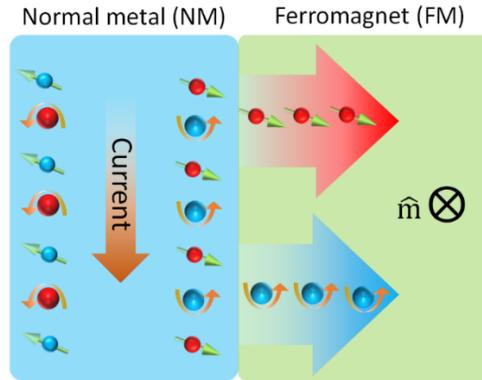

Figure 1. Schematic diagram of spin current and orbital current from non-magnetic metal to ferromagnetic layer.

The spin-pumping effect may also occur in materials with weak SOC. This letter reports on the experimental observation of strong spin pumping in a ferromagnetic thin film due to OHE in a neighboring Ti thin film with weak SOC. We modulate the OHE of Ti by adjusting the thickness, aiming to investigate the trend of spin pumping in the Ti/Co bilayers. The results indicate that the trend of damping variation with thickness perfectly coincides with the trend of accumulated orbital current in Ti layer measured by magneto-optical Kerr effect (MOKE). Building on the changes in damping, further investigation into the trend of spin-mixing conductance reveals that the spin-mixing conductance of Ti/Co, at $474.1 \times 10^{18}\ m^{-2}$, is 33.8 times higher than that of Pt/Co[], which is $14.2 \times 10^{18}\ m^{-2}$. These findings provide unprecedented opportunities for advancing the understanding of the orbital angular momentum dynamics in solids.

## 2. Results and Discussions

For the light metal Ti, there have been relevant reports indicating that the orbital Hall effect strengthens with the increasing thickness of Ti. We prepared samples ranging from 5 nanometers to 60 nanometers. The Ti film were grown in Si (111) substrates by sputtering, as described in the Supporting Information. Figure 2 presents the structural properties of Ti film. Figure 2. (a) gives a wide-angle X-ray diffraction (XRD) spectrum for a 40-nanometer thick Ti film. Figure 2. (b) shows XRD peaks for Ti films with different thicknesses. The dashed vertical lines in Figure 2. (b) indicate the expected positions for the Ti (002) and Ti (101) peaks, and the dotted indicate the Si (111) and Si (222) peaks, respectively. Figure 2. (c) shows an atomic force microscopy (AFM) surface image; the indicated roughness value is determined by averaging over AFM measurements on five different 5 μm × 5 μm areas; the uncertainty is the corresponding standard deviation.

Two results are evident from the data in Figure 2. (1) When the thickness of the Ti film is within 30 nm, the Ti (101) peak overwhelmingly dominates. However, as the thickness increases to 30 nm, the second peak Ti (002) begins to emerge. The film transitions from a single-crystal state to a polycrystalline state, and with further thickness increase, the Ti (002) peak becomes predominant. (2) The film has a very smooth surface, even when the film grows to a thickness of 40nm, the roughness is still maintained at a relatively low level about 0.89 nm, as Figure 2 (c). This smooth surface facilitates the fabrication of heterostructures with high-quality interfaces for spin-pumping studies.

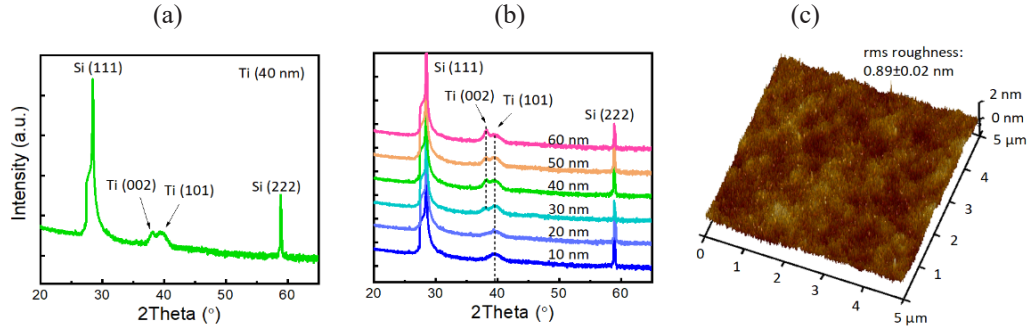

Figure 2. The properties of Ti thin films with varying thickness grown on Si (111) substrates. (a) The X-ray diffraction spectrum (XRD) of a 40-nanometer thick Ti film. (b) Comparison of XRD data for Ti thin films with different thicknesses. (c) Atomic force microscopy surface image of a 40 nm thick Ti film.

Figure 3. shows the Ferromagnetic resonance (FMR) measured on an Si/Co (6 nm) sample and an Si/Ti (40 nm)/Co (6 nm) sample at 12 GHz. It is essential to emphasize that in all comparison samples in this paper, the thickness of Co is 6 nm, and the growth processes are identical. Figure 3 (a) provides a schematic representation of the spin pumping process, where the spin current is converted into an orbital current within the magnetic layer. Then, the orbital current is resonantly absorbed in the non-magnetic metal. Figures 3. (b) and (c) demonstrate the peak-to-peak FMR linewidth (Δ$H$pp) values, the linewidth of Si/Ti (40 nm)/Co (6 nm) is significantly broader, almost 4.34 times greater than that of Si/Co (6 nm). Further, Figure 3. (c) shows that the Lorentzian fit is much better than the Gaussian fit, indicating that the film inhomogeneity

contribution to $\Delta H$pp is relatively small. These results together suggest that a significant damping due to spin pumping is present in the sample with a Ti layer, as discussed shortly.

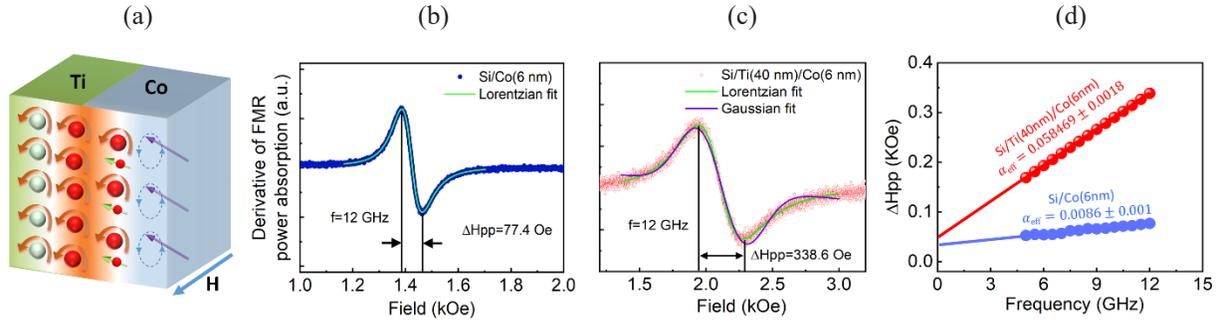

Figure 3. OAM-enhanced spin pumping in Ti/Co structures. (a) Conceptual diagram. (b) FMR profile of Si/Co(6 nm). (c) FMR profile of Si/Ti(40 nm)/Co(6 nm). d) FMR linewidth as a function of frequency for the two samples.

To confirm the above-discussed, Ti-produced linewidth broadening, FMR measurements were repeated at different frequencies (f). The resulting $\Delta H$pp versus f responses are shown in Figure 3 (d) One can see that $\Delta H$pp in the sample with an Ti (40nm) layer is larger at all frequencies, confirming the result shown in Figure 3 (b) and (c). The lines in Figure 3 (d) show fits to

$$\Delta H\text{pp} = \frac{2\alpha_{\text{eff}}}{\sqrt{3}|\gamma|}f + \Delta H_0 \qquad (1)$$

where $\alpha_{\text{eff}}$ is the effective damping constant, $|\gamma|$= 2.2 MHz/Oe is the absolute gyromagnetic ratio, and $\Delta H_0$ denotes the inhomogeneity-caused line broadening[]. The fitting-yielded $\alpha_{\text{eff}}$ values are indicated in the figure. One can see that $\alpha_{\text{eff}}$ in the sample with the Ti (40 nm) film is about 670% larger than that in the one without an Ti (40 nm) film. In addition to the previously mentioned point that the Lorentzian fitting being superior to Gaussian fitting demonstrates the good uniformity of the film, there are two further aspects here that can further prove that the increase in damping is not due to the inhomogeneity of the film. First, the $\Delta H_0$ (38.3 Oe) value is much smaller than the linewidth (338.6 Oe), indicating good film uniformity. Additionally, the fitting values of $\Delta H_0$ for samples Si/Ti (40 nm)/Co (6 nm) and Si/Co (6 nm) are extremely close, suggesting that the inserted Ti (40 nm) layer has a very small impact on the uniformity of Co.

Based on the experimental foundation mentioned above, a series of Ti($t_{\text{Ti}}$)/Co(6 nm) samples were prepared, where $t_{\text{Ti}}$ represents the thickness of Ti, ranging from 5 nm to 60 nm. We measured the linewidth versus frequency data for each sample, and a portion of the obtained data is shown in Figure 4(a). With the increase in Ti thickness, the slope of $\Delta H$pp versus f increased when fitted using Equation (1). Within the range of $t_{Ti}$ up to 60 nm, $\Delta H_0$ showed almost no significant change, as shown in Figure 4(b). This implies that the increase in Ti thickness does not markedly alter the uniformity of the Co layer. However, when growing a thicker Ti layer (> 60 nm), $\Delta H_0$ sharply increases. This is because maintaining a smooth surface becomes challenging for thicker films, leading to pronounced inhomogeneity in the deposited Co film, as detailed in the Supporting Information. So the primary focus of this paper is limited to Ti thicknesses less than or equal to 60 nm, the obtained damping trend with thickness variation is shown in Figure 4 (c). To extract the OHE contribution to damping ($\alpha_{eff}$) and determine orbital relaxation length ($l_L$), we base on the relationship between the orbital torque effective ($\xi_{DL}$) and $t_{Ti}$:

$$\xi_{DL} = C_{FM}\theta_{OH}\left(1 - \text{sech}\left(\frac{t_{Ti}}{l_L}\right)\right) + \xi_0 \qquad (2)$$

Where $\theta_{OH}$ refers to the orbital Hall angle, $C_{FM}$ represents the conversion efficiency from orbital current to spin current in the magnetic layer and $\xi_0$ is the $t_{Ti}$-independent part of $\xi_{DL}$, which may arise from interfacial origins such as interfacial Rashba effect or interfacial spin-orbit torque. As in the qualitative analysis of Figure 3 (a), the increase in OHE leads to an enhanced efficiency of resonant absorption. This, in turn, results in an increased damping of the magnetic layer. Therefore, the relationship can be expressed as $\xi_{DL} \propto \alpha_{eff}$, we examine the increasing trend of $\alpha_{eff}$ with $t_{Ti}$, the expression can that can be written by

$$\alpha_{eff} \propto \left(1 - \text{sech}\left(\frac{t_{Ti}}{l_L}\right)\right) + \alpha_0 \qquad (3)$$

Where $\alpha_0$ refers to the damping constant of the 6nm thick single-layer Co. The contribution of the Ti layer to damping ($\alpha_{Ti}$) comes from $\left(1 - \text{sech}\left(\frac{t_{Ti}}{l_L}\right)\right)$. We fitted Figure 4(c) with Equation (3) and obtained an orbital coherence length ($l_L$) of 51.12±7.735 nm, which is in good agreement with the length value of 50±15 nm reported based on Equation (2). However, further investigation reveals that $t_{Ti}$ increases from 5 nm to 60 nm, damping increases from 0.007 to 0.094, nearly a 13.4-time increase, which coincides with the observed trend in the accumulation of orbital current in a single-layer Ti, as observed by the MOKE method.

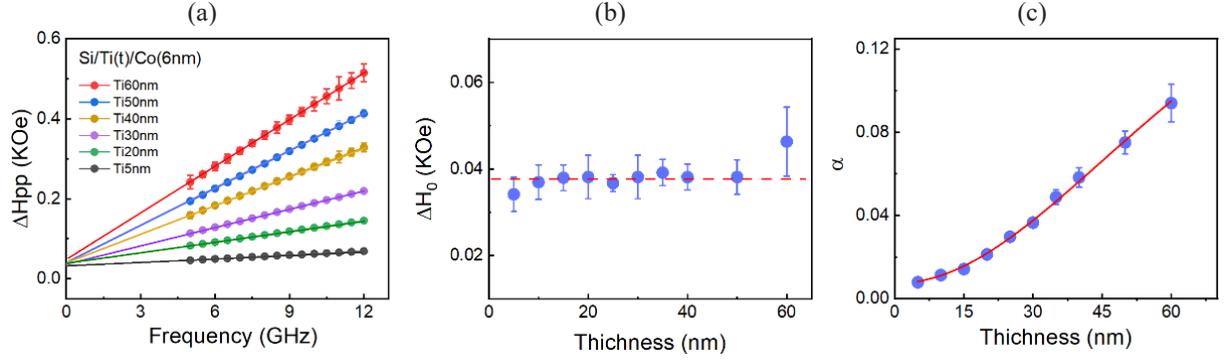

Figure 4. OHE is positively correlated with damping. (a) Comparison of FMR linewidth versus frequency responses measured on different thicknesses. (b) Inhomogeneity-caused line broadening versus thicknesses on different samples. (c) Damping versus thicknesses on different samples, the blue dots represent the data points, and the solid red line represents the fitting line.

Based on the trend of damping with different thicknesses of Ti, evaluate the spin-mixing conductance as shown in Equation (4), where $Re(g_{\uparrow\downarrow}^{eff})$ is the real part of the spin mixing conductance, $d$ represents the thickness of Ti, $\alpha_{Co}$ represents the damping of a 6nm thick Co, and $\alpha_{Ti/Co}$ represents the damping of the Ti($d$)/Co(6 nm) structure.

$$Re(g_{\uparrow\downarrow}^{eff}) = \frac{4\pi M_s d}{\hbar|\gamma|}(\alpha_{Ti/Co} - \alpha_{Co}) \qquad (4)$$

As shown in Figure 5, with the increase in Ti thickness, the spin-mixing conductance gradually increases. Particularly, when the thickness of Ti reaches 60 nm, the spin-mixing conductance is as high as $474.1 \pm 3 \times 10^{18} \, m^{-2}$. A comparison for Ti/Co, Pt/Co, W/Co, Ta/Co, and Ir/Co reveals that Ti/Co is an order of magnitude higher than the heavy metal structures. If we continue to compare with materials having stronger spin-orbit coupling, such as the topological materials $BiSbTe_{1.5}Se_{1.5}$ (BSTS)/Cu/CoFeB and $(Bi, Sb)_2Te_3$ (BST)/Ru/CoFeB, as well as the topological Dirac semimetal α-Sn/Ag/Py, we observed that the spin-mixing conductance of the orbital current is significantly higher than that of the spin current. This suggests that the orbital current is less affected by the interface, making it more conducive to transmission at the interface.

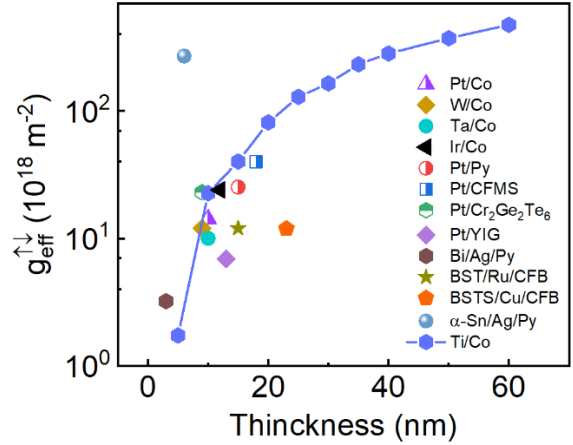

Figure 5. Comparison of spin-mixing conductance for Ti/Co and different materials.

## 3. Conclusions

We investigated the damping and spin-mixing conductance of Ti/Co based on the spin pumping effect. As the thickness of Ti increases, the changes in damping and spin-mixing conductance align perfectly with the conclusions drawn from previous MOKE measurements of orbital current accumulation. The significant spin-mixing conductance indicates that the orbital current is less affected by the interface, making it more easily transmissible across the interface. This not only provides insights for optimizing and improving orbital torque devices in the future but also serves as a reference for enhancing spin-orbit torque efficiency in the future.

**Availability of data and materials**

All data needed to evaluate the conclusions in the paper are present in the paper and/or the Supplementary Materials.

**Financial support and sponsorship**

We thank the National Natural Science Foundation of China (No. 92271120).

**Conflicts of interest**

All authors declared that there are no conflicts of interest.

**Ethical approval and consent to participate**

Not applicable.

**Consent for publication**

Not applicable.

Supporting Information

# Gigantic spin mixing conductance caused by the orbital hall effect

**S1. The details of thin film growth by sputtering**

Ti films were grown on Si (111) substrates using the direct current (DC) sputtering mode. The main steps are as follows: (1) Substrate cleaning: Si substrates were soaked in acetone and subjected to 5 minutes of ultrasonic cleaning, followed by another 5 minutes of cleaning in alcohol, and finally rinsed with deionized water. (2) Pre-growth preparation: After drying the substrates, they were transferred to a pre-vacuum chamber. The vacuum level of the magnetron chamber was ensured to be as low as $5*10^{-9}$ Torr, then the substrate was transferred to the sputtering chamber. And prior to film growth, the substrate was heated to 200 degrees Celsius and maintained for 20 minutes. After stopping the heating process and allowing the chamber to return to room temperature, and film growth commenced once the chamber reached a vacuum level of $5*10^{-9}$ torr again. (3) Film growth: With the substrate rotating at 30 revolutions per minute and argon pressure stabilized at 3 mtorr, Ti was pre-sputtered off-axis target for 5 minutes at a power of 35W. Subsequently, the sputtering power was reduced to 30W, and the shutter was opened to initiate film growth. The growth rate of Ti was 0.8 nm/min.

**S2. Surface morphology**

We prepared Ti films with different thicknesses, and as the thickness increased, the surface roughness began to rise. The morphology for partial thicknesses are shown in the Figure S1, where (a), (b), (c), (d), (e), and (f) represent thicknesses of 5 nm, 10 nm, 20 nm, 30 nm, 40 nm, and 60 nm, respectively. The corresponding roughness values are 0.05±0.01 nm, 0.11±0.03 nm, 0.39±0.01 nm, 0.75±0.04 nm, 0.89±0.02 nm, and 2.35±0.04 nm. Particularly, when the thickness reached 60 nm, the roughness increased abruptly from 0.91±0.04 nm at 50nm to 2.53±0.04 nm at 60 nm. If the thickness continues to increase, it will lead to more severe roughness issues, affecting the uniformity of the subsequently deposited Co film. Therefore, to ensure the uniformity of the film is not restricted by roughness, we strictly limit the roughness to 1 nm as the effective research range. So, this study focuses on results for Ti film thicknesses less than 60 nm.

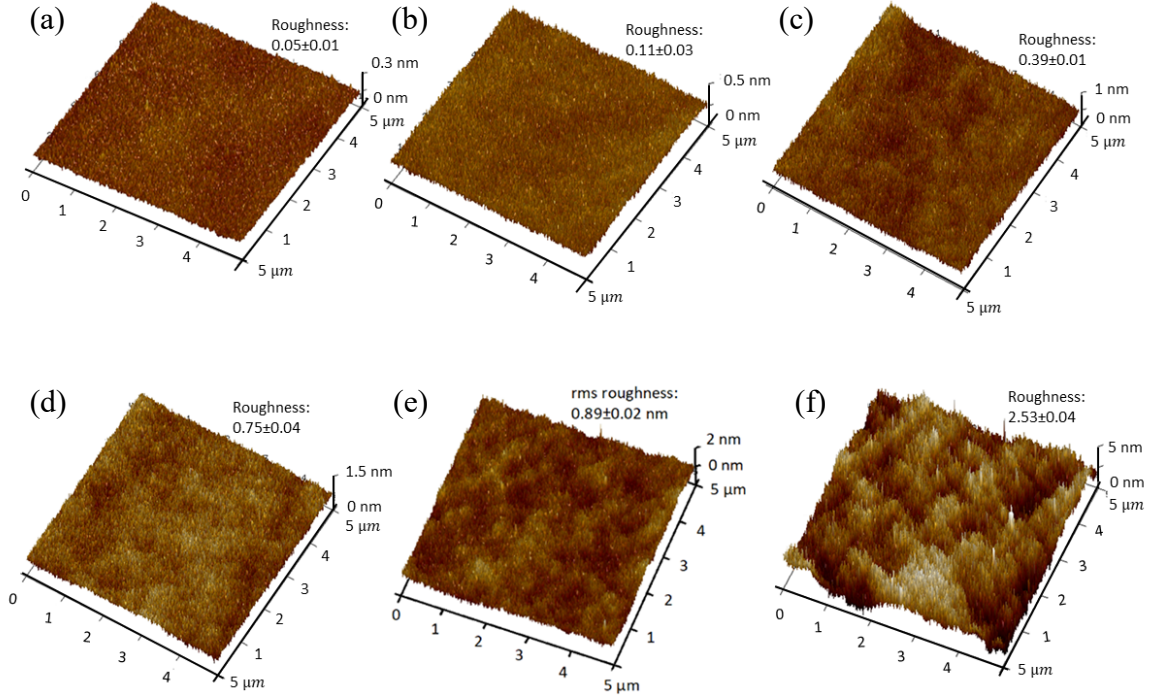

Figure S1. Surface morphology of Ti with different thicknesses. where (a), (b), (c), (d), (e), and (f) represent thicknesses of 5 nm, 10 nm, 20 nm, 30 nm, 40 nm, and 60 nm, respectively. The corresponding roughness values are 0.05±0.01 nm, 0.11±0.03 nm, 0.39±0.01 nm, 0.75±0.04 nm, 0.89±0.02 nm, and 2.35±0.04 nm.

## S3. Ferromagnetic Resonance Fields and Anisotropy

By performing Lorentz fits to the main text, the relationship between the resonance field ($H_{FMR}$) and frequency (f) can also be obtained, as indicated by the Kittel equation:

$$f = |\gamma|\sqrt{H_{FMR}(H_{FMR} + 4\pi M_{eff})} \qquad (S1)$$

where $4\pi M_{eff}$ denotes the difference between the saturation induction $4\pi M_s$ and the effective anisotropy field $H_a$ of the Co film, namely, $4\pi M_{eff} = 4\pi M_s - H_a$. Note that $H_a > 0$ and $H_a < 0$ correspond to the presence of a perpendicular uniaxial anisotropy and an easy-plane anisotropy, respectively. The fitting-produced $4\pi M_{eff}$ values are given in Figure S3. one can evaluate $H_a = 4\pi M_s - 4\pi M_{eff}$. The obtained $H_a$ values are less than 3% of $4\pi M_s$. These values may suggest the presence of anisotropy in the Co film, but may also result from the $4\pi M_s$ errors due to, for example, errors in estimating the volumes of the Co films.

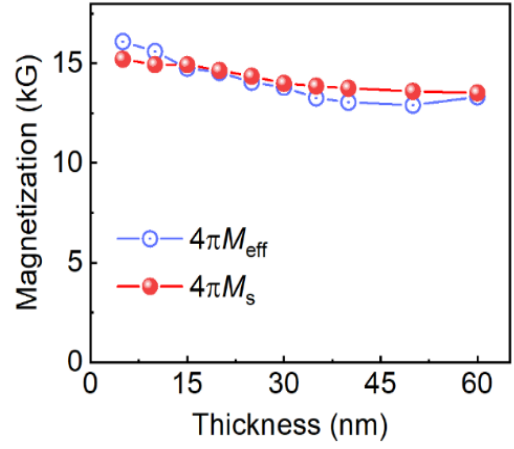

Figure S2. Comparison of the $4\pi M_{eff}$ values from FMR measurements and the $4\pi M_s$ values from VSM measurements in Ti(t)/Co, t is the thickness.

## S4. Comparison of spin-mixing conductance

Main text Figure 5 compares the spin mixing conductance of different structures. However, it lacks a corresponding comparison of the spin diffusion lengths (λ) and the test environment temperatures. Table 1 presents these three indicators together as follows. Through the spin diffusion lengths, it is observed that the diffusion distance of spin currents in strong spin-orbit coupling materials is generally shorter, while the diffusion length of orbital currents is nearly ten times that of spin currents, indicating that orbital currents are more suitable for long-distance transmission in crystals.

Table 1. Comparison of spin-mixing conductance for different material structures.

|  | Temperature (K) | $g_{\text{eff}}^{\uparrow\downarrow}(10^{18}\,m^{-2})$ | $\lambda_d(nm)$ | ref |
|---|---|---|---|---|
| Au/Co | 300 | 10.38 ± 1 | 8.3 ± 0.4 | [] |
| W/Co | 300 | 12 ± 1 | 6.1 ± 0.4 | [] |
| Ta/Co | 300 | 10 ± 1 | 5.5 ± 0.4 | [] |
| Ir/Co | 300 | 24 ± 2 | 6.3 ± 0.4 | [] |
| Pt/Co | 300 | 14.2 ± 2 | 8 ± 0.4 | [] |
| Pt/Py | 300 | 25.3 ± 5 | 8.4 ± 0.3 | [] |
| Pt/CFMS | 300 | 40 ± 5 | 7.5 ± 0.3 | [] |
| Pt/Cr$_2$Ge$_2$Te$_6$ | 10 | 23 ± 5 | 14 | [] |
| Pt/YIG | 300 | 6.9 ± 0.6 | 7.3 ± 0.2 | [] |
| Bi/Ag(t)/Py | 300 | 3.21 | - | [] |
| BST/Ru(or Ti)/CoFeB | 300 | 12 ± 1 | - | [] |
| BSTS/Cu/CoFeB/SiO$_2$ | 300 | 11.9 ± 0.34 | - | [] |
| α-Sn/Ag/Py | 300 | 270 ± 2 | - | [] |
| Ti/Co | 300 | 474.1 ± 3 | 51.12 ± 7.7 | This work |